\documentclass[twoside]{dis09}
\usepackage[latin1]{inputenc}
\usepackage[dvips]{graphicx,epsfig,color}
\usepackage{wrapfig,rotating}
\usepackage{amssymb,amsmath,array}

\input{paperdef}

\begin{document}

\thispagestyle{empty}
\setcounter{page}{0}
\def\thefootnote{\fnsymbol{footnote}}

\begin{flushright}
\mbox{}
arXiv:0906.4677 [hep-ph]
\end{flushright}

\vspace{1cm}

\begin{center}

{\large\sc {\bf Indirect Limits on Higgs and SUSY Masses}}
\footnote{invited talk given at the {\em DIS09}, 
April 2009, Madrid, Spain}

\vspace{1cm}

{\sc 

S.~Heinemeyer%
\footnote{
email: Sven.Heinemeyer@cern.ch}
}

\vspace*{1cm}

{\it
Instituto de F\'isica de Cantabria (CSIC-UC), 
Santander,  Spain
}
\end{center}

\vspace*{0.2cm}

\BC {\bf Abstract} \EC
We review the current best fit values of Higgs boson masses in the
Standard Model (SM) and its minimal supersymmetric extension (MSSM)
obtained from existing experimental data. 
We also review the parameters space of the constrained MSSM (CMSSM) and
the non-universal Higgs mass model (NUHM1) currently preferred by
precision data. Following a Frequentist approach,
the experimental data includes electroweak precision
observables, $B$~physics observables and the relic density of cold dark
matter.

\def\thefootnote{\arabic{footnote}}
\setcounter{footnote}{0}

\newpage


\hyphenation{re-commend-ed Post-Script}

\graphicspath{{figs/}}

\title{Indirect Limits on Higgs and SUSY Masses}

\author{Sven Heinemeyer
%
%
\vspace{.3cm}\\
%
Instituto de F\'isica de Cantabria (CSIC), 39005 Santander, Spain
}

\maketitle

\begin{abstract}
We review the current best fit values of Higgs boson masses in the
Standard Model (SM) and its minimal supersymmetric extension (MSSM)
obtained from existing experimental data. 
We also review the parameters space of the constrained MSSM (CMSSM) and
the non-universal Higgs mass model (NUHM1) currently preferred by
precision data. Following a Frequentist approach,
the experimental data includes electroweak precision
observables, $B$~physics observables and the relic density of cold dark
matter. 

\end{abstract}


\section{Introduction}

Identifying the mechanism of electroweak symmetry
breaking will be one of the main goals of the LHC. 
Many possibilities have been studied in the literature, of which 
the most popular ones are the Higgs mechanism within the Standard Model
(SM)~\cite{sm} and within the Minimal Supersymmetric Standard Model
(MSSM)~\cite{mssm}. 

Theories based on Supersymmetry (SUSY)~\cite{mssm} are widely
considered as the theoretically most appealing extension of the
SM. They are consistent with the approximate
unification of the gauge coupling constants at the GUT scale and
provide a way to cancel the quadratic divergences in the Higgs sector
hence stabilizing the huge hierarchy between the GUT and the Fermi
scales. Furthermore, in SUSY theories the breaking of the electroweak
symmetry is naturally induced at the Fermi scale, and the lightest
supersymmetric particle can be neutral, weakly interacting and
absolutely stable, providing therefore a natural solution for the dark
matter problem.
SUSY predicts the existence of scalar partners $\tilde{f}_L,
\tilde{f}_R$ to each SM chiral fermion, and spin--1/2 partners to the
gauge bosons and to the scalar Higgs bosons. 
The Higgs sector of the Minimal Supersymmetric Standard Model (MSSM) 
with two scalar doublets accommodates five physical Higgs bosons. In
lowest order these are the light and heavy $\cp$-even $h$
and $H$, the $\cp$-odd $A$, and the charged Higgs bosons $H^\pm$.
Higher-order contributions yield large corrections to the masses
and couplings~\cite{habilSH,mhiggsAWB}

So far, the direct search for SUSY particles has not been successful.
One can only set lower bounds of ${\cal O}(100)$~GeV on
their masses~\cite{pdg}. The search reach will be extended in various
ways in the ongoing Run~II at the upgraded Fermilab
Tevatron~\cite{TevFuture}. 
The LHC~\cite{atlas,cms} and the $e^+e^-$ International Linear Collider
(ILC)~\cite{ilc} have very good prospects for
exploring SUSY at the TeV scale, which is favored from naturalness
arguments. From the interplay of both machines detailed
information on many SUSY can be expected in this case~\cite{lhcilc}.

Besides the direct detection of SUSY particles (and Higgs bosons), 
physics beyond the SM can also be probed by precision observables via the
virtual effects of the additional particles.
Observables (such as particle masses, mixing angles, asymmetries etc.)
that can be predicted within a certain model and thus depend sensitively
on the other model parameters constitute a test of the model on the
quantum level. Various models predict different values of the same
observable due to their different particle content and
interactions. This permits to distinguish between, for instance, the SM and 
the MSSM via precision observables.


\section{Higgs mass predictions in the SM}

Within the SM the last unknown parameter, the mass of the Higgs boson
$\MHSM$, can be predicted as described above. The fit is based on the
electroweak precision observables (EWPO) measured at LEP and
SLD and the Tevatron~\cite{lepewwg,LEPEWWG,TEVEWWG} and can include or
exclude the direct searches 
performed at LEP~\cite{LEPHiggsSM} and at the Tevatron~\cite{TevHiggsSM}.

\begin{figure}[htb!]
\begin{center}
\includegraphics[width=.48\textwidth,height=.50\textwidth] 
                {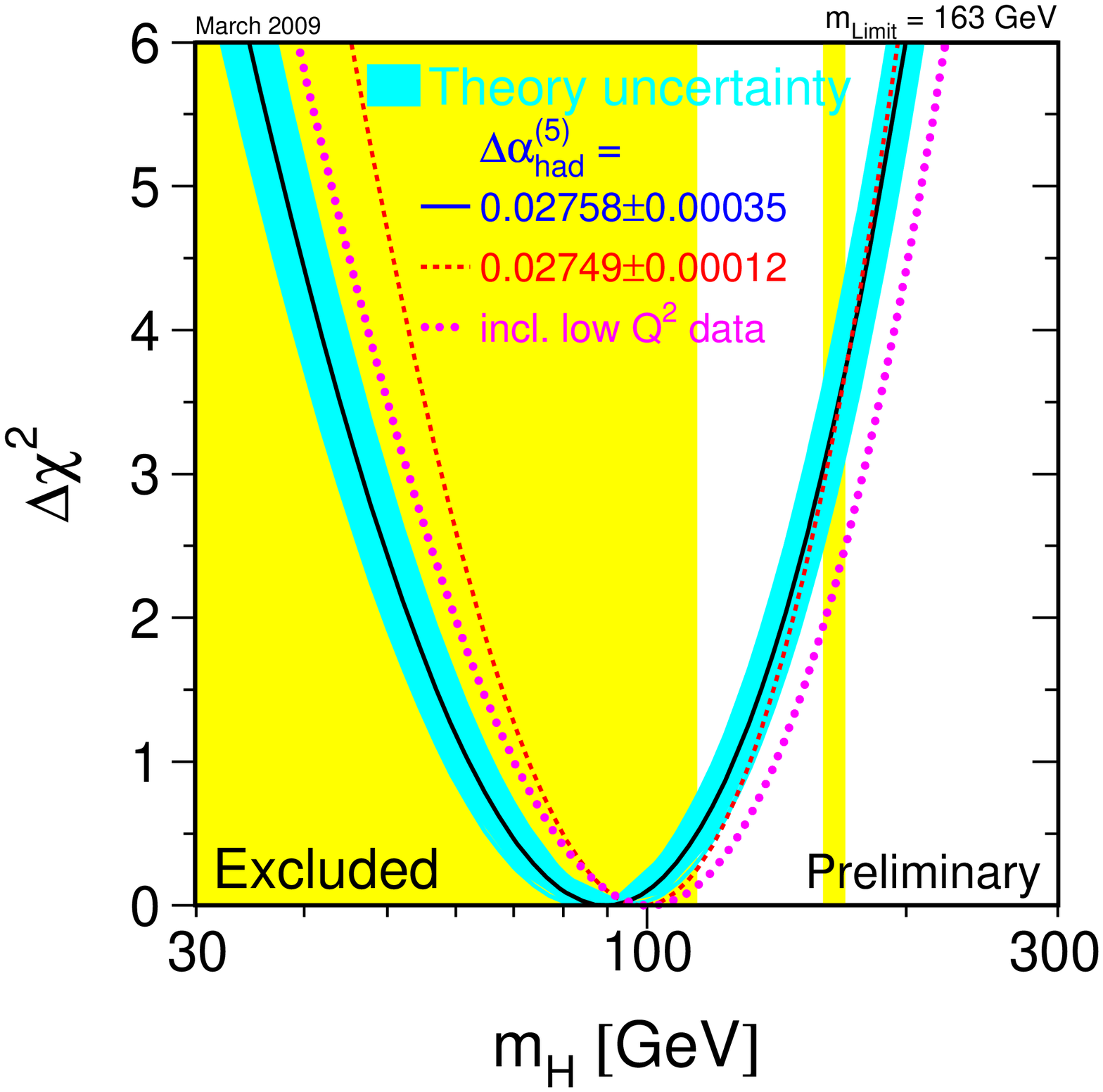}
\hspace{2mm}
\includegraphics[width=.48\textwidth,height=.47\textwidth] 
                {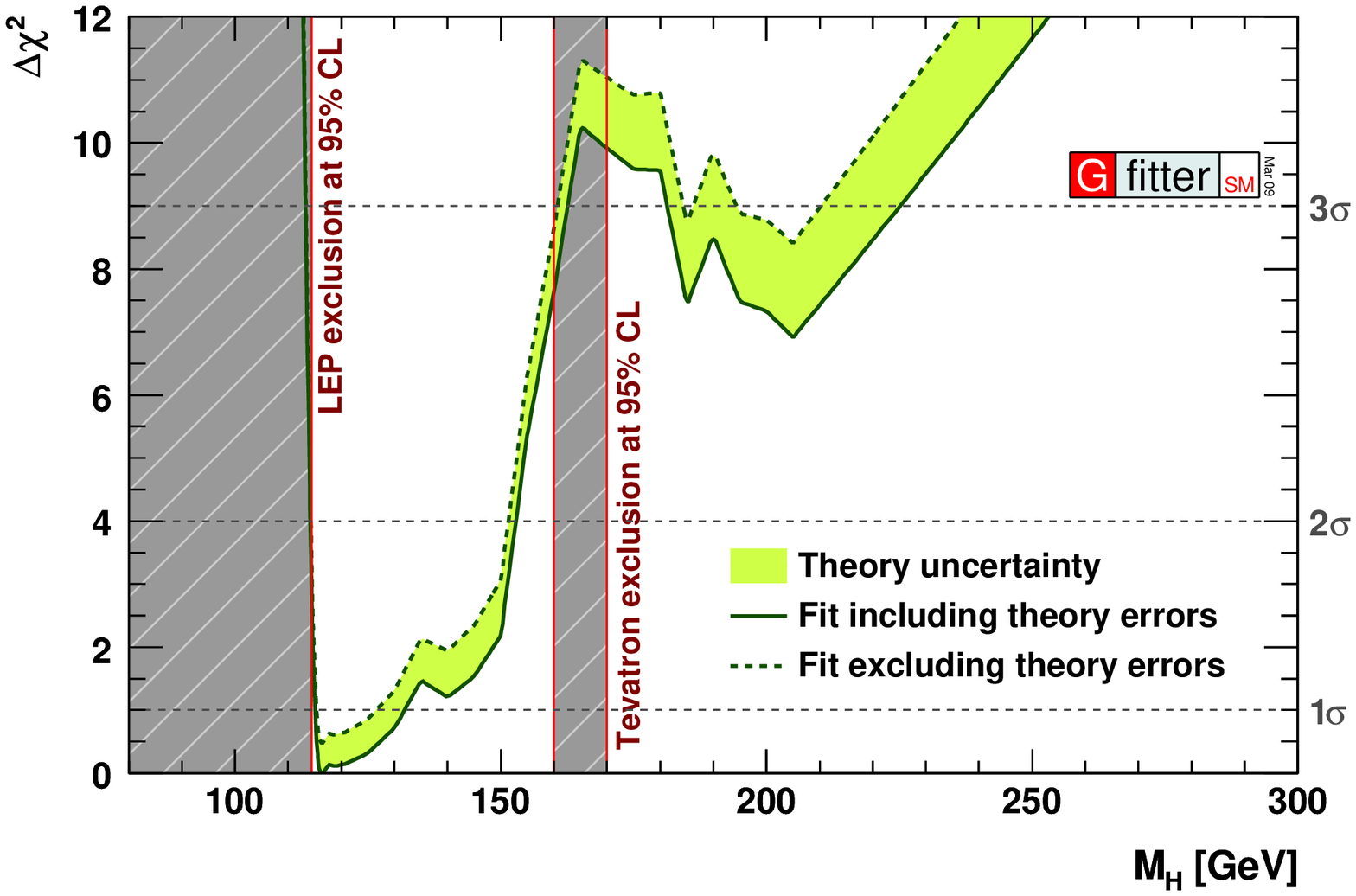}
\caption{%
$\De\chi^2$ curve derived from all EWPO measured at LEP, SLD, CDF and
D0, as a function of $\MHSM$, assuming the SM to be the correct theory
of nature. The results of the direct searches are (not) included in the
right~\cite{GFitter} (left~\cite{LEPEWWG}) plot. 
}
\label{fig:MhSM}
\end{center}
\end{figure}

In \reffi{fig:MhSM} we show the result for
the global fit to $\MHSM$ based on all EWPO. 
The ``blue band plot''~\cite{LEPEWWG} shown on the left side of
\reffi{fig:MhSM} excludes the direct searches, the right
plot~\cite{GFitter} includes these searches. In both plots 
$\De\chi^2$ is shown as a function of $\MHSM$. Excluding the direct
searches yields%
\footnote{A slightly tighter bound can be expected once all 
experimental results for $\MW$ will have been combined.}%
\begin{align}
\MHSM = 90^{+36}_{-29} \gev~, \quad \MHSM \le 163 \gev ~(95\% \mbox{~C.L.}),
\label{MHSMfit}
\end{align}
still compatible with the direct LEP bound of~\cite{LEPHiggsSM}
\begin{align}
\MHSM \ge 114.4 \gev ~(95\% \mbox{~C.L.})
\label{MHSMlimit}
\end{align}
The
theory (intrinsic) uncertainty in the SM calculations (as evaluated with 
{\tt TOPAZ0}~\cite{topaz0} and {\tt ZFITTER}~\cite{zfitter}) are
represented by the thickness of the blue band. The width of the parabola
itself, on the other hand, is determined by the experimental precision of
the measurements of the EWPO and the SM input parameters.

If the direct searches are included, shown in the right plot
of \reffi{fig:MhSM}, the preferred region changes
to~\cite{GFitter} 
\begin{align}
\MHSM = 116.4^{+18.3}_{-1.4} \gev~, \quad 
\MHSM \le 152 \gev ~(95\% \mbox{~C.L.})
\label{MHSMfit2}
\end{align}


\section{Higgs mass predictions in the CMSSM}

A fit as close as possible to the SM fit for $\MHSM$ (resulting in
the left plot of \reffi{fig:MhSM}) has been performed in
\citere{mastercode1} for the lightest Higgs boson in the Constrained
MSSM (CMSSM).
All EWPO as in the SM~\cite{LEPEWWG} (except $\Ga_W$, which has a
minor impact) were included, supplemented by the Cold Dark Matter
constraint, the $(g-2)_\mu$ results and the
$\br(b \to s \ga)$ constraint (see \citeres{mastercode1,mastercode2} for
details and a complete list of references).
Following a Frequentist approach, 
the $\chi^2$ is minimized with respect to all CMSSM parameters for
each point of this scan. Therefore, $\De \chi^2=1$ represents the
68\% confidence level uncertainty on $\Mh$.
Since the direct Higgs boson search limit from LEP is not used in this
scan the lower bound on $\Mh$ arises as a consequence of 
{\em indirect} constraints only, as in the SM fit.

\begin{wrapfigure}{r}{0.5\columnwidth}
\vspace{2em}
\begin{picture}(500,190) 
  \put(0,10){ \resizebox{7.5cm}{!}
             {\includegraphics[angle=0]{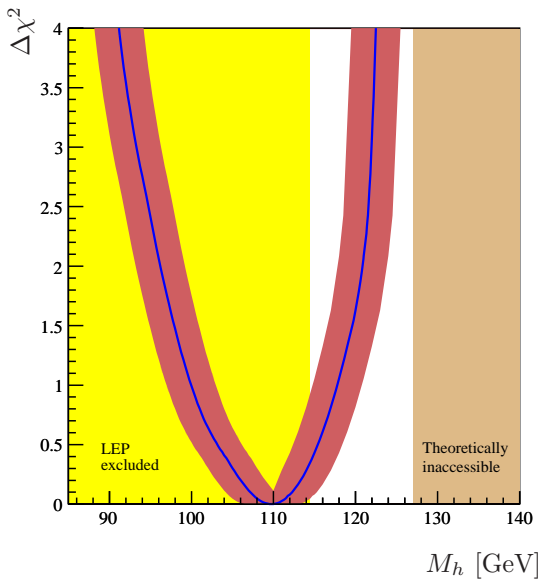}}}
  \put(160, 05){$\Mh$ [GeV]}
  \put(10, 200){\begin{rotate}{90}$\Delta \chi^2$\end{rotate}}
\end{picture}
\vspace{-2.5em}
\caption{Scan of the lightest Higgs boson mass versus $\De\chi^2$ in the
  CMSSM~\cite{mastercode1} (using $\mt = 170.9 \pm 1.8 \gev$). 
  The direct limit
  on $\Mh$ from LEP~\cite{LEPHiggsSM,LEPHiggsMSSM} is not included.
  The red (dark gray) band represents the total
  theoretical uncertainty from unknown higher-order corrections.}
\label{fig:MhCMSSM}
\end{wrapfigure}

In \reffi{fig:MhCMSSM}~\cite{mastercode1} 
$\De\chi^2$ is shown as a function of $\Mh$ in the CMSSM. The area
with $\Mh \ge 127$ is theoretically inaccessible. There is a
well defined minimum in the red band parabola, leading to a prediction
of~\cite{mastercode1}  
\begin{align}
\Mh^{\rm CMSSM} = 110^{+8}_{-10}\;{\small{\rm(exp)}} 
                     \pm 3\;{\small{\rm(theo)}\gev ,}
\label{MH_CMSSM}
\end{align}
where the first, asymmetric uncertainties are experimental and the
second uncertainty is theoretical (from the unknown higher-order
corrections to $\Mh$~\cite{mhiggsAEC,PomssmRep}). 
(An update using the latest $\mt$ measurements is currently
being prepared~\cite{mastercode3}.) 
The fact that the minimum in \reffi{fig:MhCMSSM} is
sharply defined is a general consequence of the MSSM, where the
neutral Higgs boson mass is not a free parameter.
The theoretical upper bound $\Mh \lsim 135 (127) \gev$ in the (C)MSSM
explains the sharper rise of the $\De\chi^2$ at large $\Mh$
values and the asymmetric uncertainty. In the SM, $\MHSM$ is a
free parameter and only enters (at leading order) logarithmically in
the prediction of the precision observables. In the (C)MSSM this
logarithmic dependence is still present, but in addition $\Mh$
depends on $\mt$ and the SUSY parameters, mainly from the scalar
top sector. The low-energy SUSY parameters in turn are all connected
via RGEs to the GUT scale parameters.  
The sensitivity on $\Mh$ is therefore the combination of the indirect
constraints on the four free CMSSM 
parameters and the fact that $\Mh$ is directly predicted in
terms of these parameters.


\section{SUSY mass predictions in the CMSSM and NUHM1}

In a similar way to the CMSSM Higgs boson mass determination also the
SUSY parameters themselves can be fitted~\cite{mastercode2}.
Many analyses in this direction have been performed, see
\citere{mastercode2} for a list of references.

In \reffi{fig:m0-m12}~\cite{mastercode2} we show the preferred regions
in the CMSSM (left) and the non-universal Higgs mass model (NUHM1)
(right) in the $m_0$-$m_{1/2}$~plane. 
The solid (dot-dashed/dashed) line shows the
regions that can be covered at CMS with 1~\ifb at $14 \tev$
(100~\ipb at $14 \tev$/50~\ipb at $10 \tev$) of {\em understood} data. It can
be seen that the LHC has good chances to discover the CMSSM or NUHM1
with early data.
The mass spectrum of the two models for the two best-fit points is shown
in \reffi{fig:bestfit}. If one of these two points were realized in
nature the LHC and the ILC could observe many SUSY particles and measure
their properties~\cite{atlas,cms,ilc}.

\begin{figure}[htb!]
\begin{center}
\includegraphics[width=.48\textwidth,height=.37\textwidth] 
                {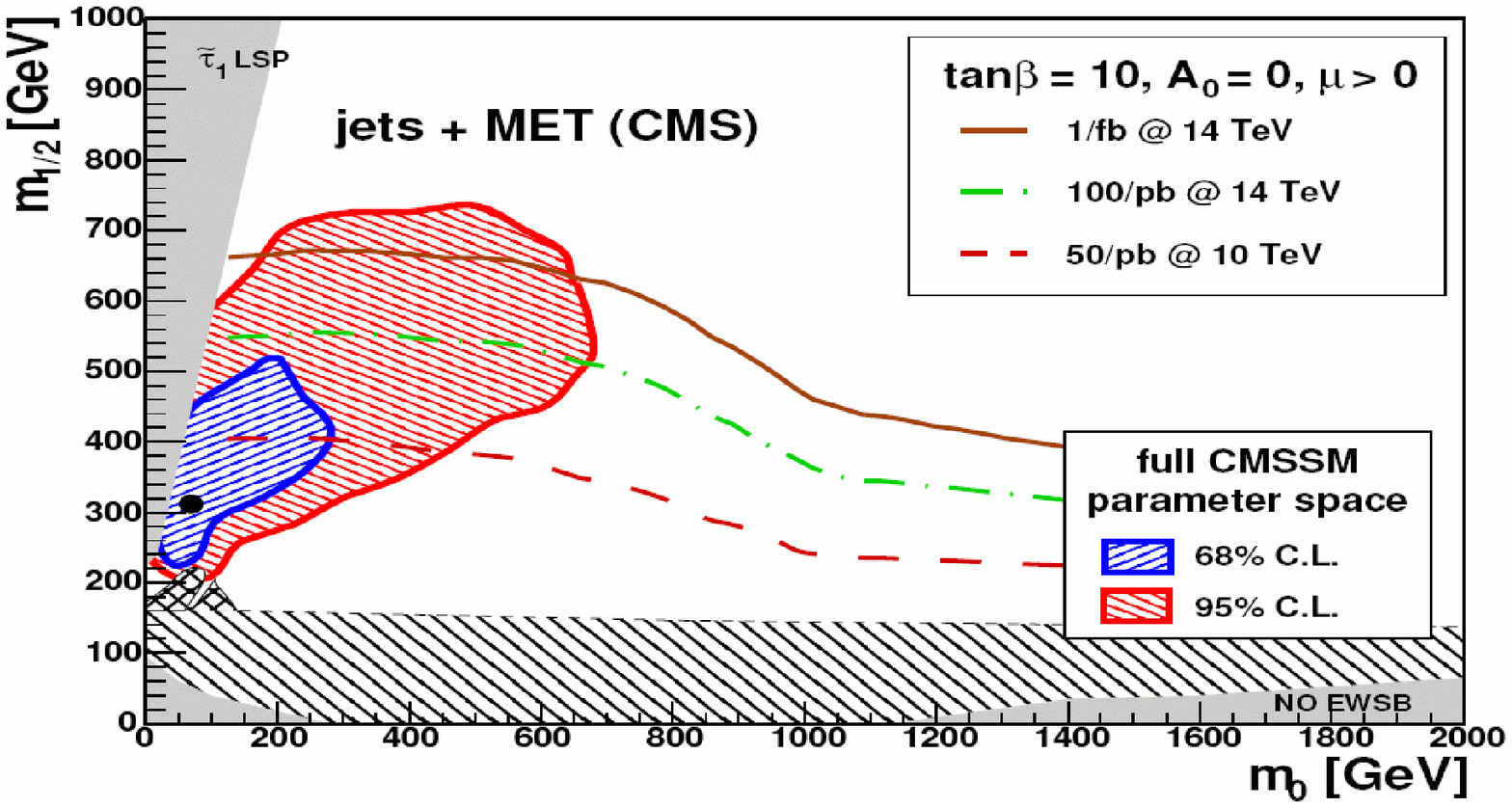}
\hspace{2mm}
\includegraphics[width=.48\textwidth,height=.37\textwidth] 
                {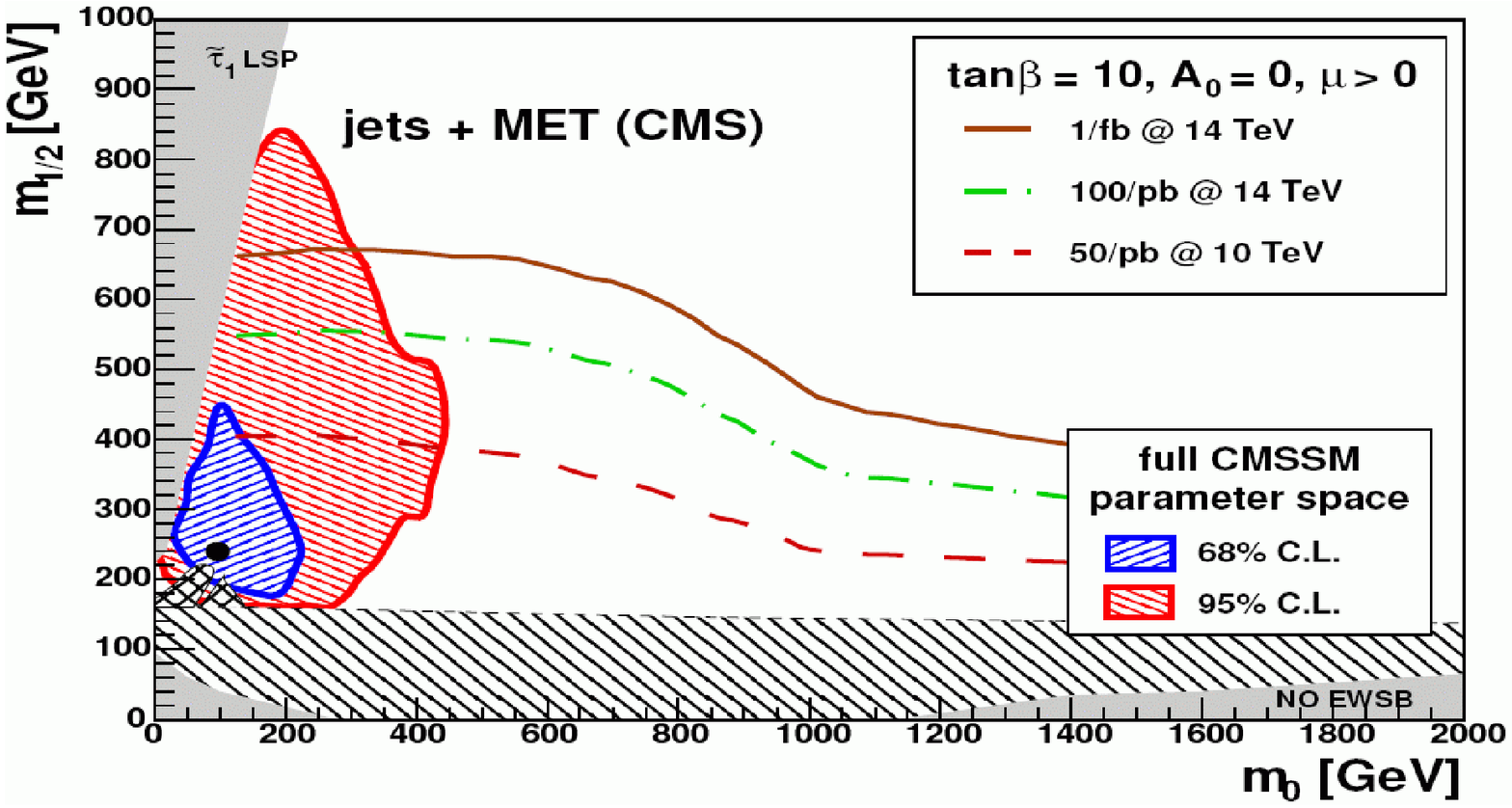}
\caption{Areas in the $m_0$-$m_{1/2}$ planes preferred by current
experimental data~\cite{mastercode2}; left: CMSSM, right: NUHM1.}
\label{fig:m0-m12}
\end{center}
\vspace{-1em}
\end{figure}

\begin{figure}[htb!]
\begin{center}
\includegraphics[width=.48\textwidth,height=.37\textwidth] 
                {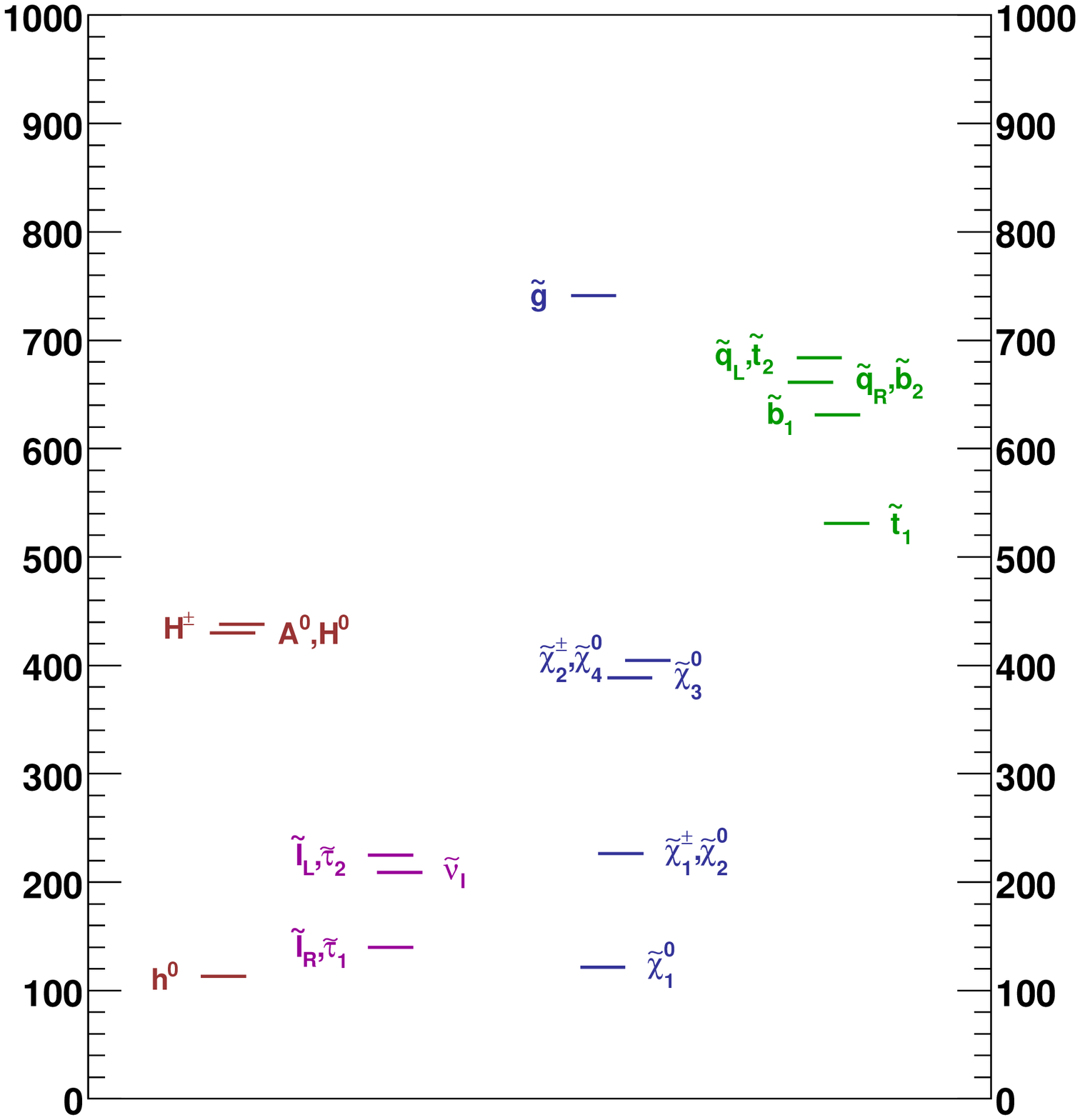}
\hspace{2mm}
\includegraphics[width=.48\textwidth,height=.37\textwidth] 
                {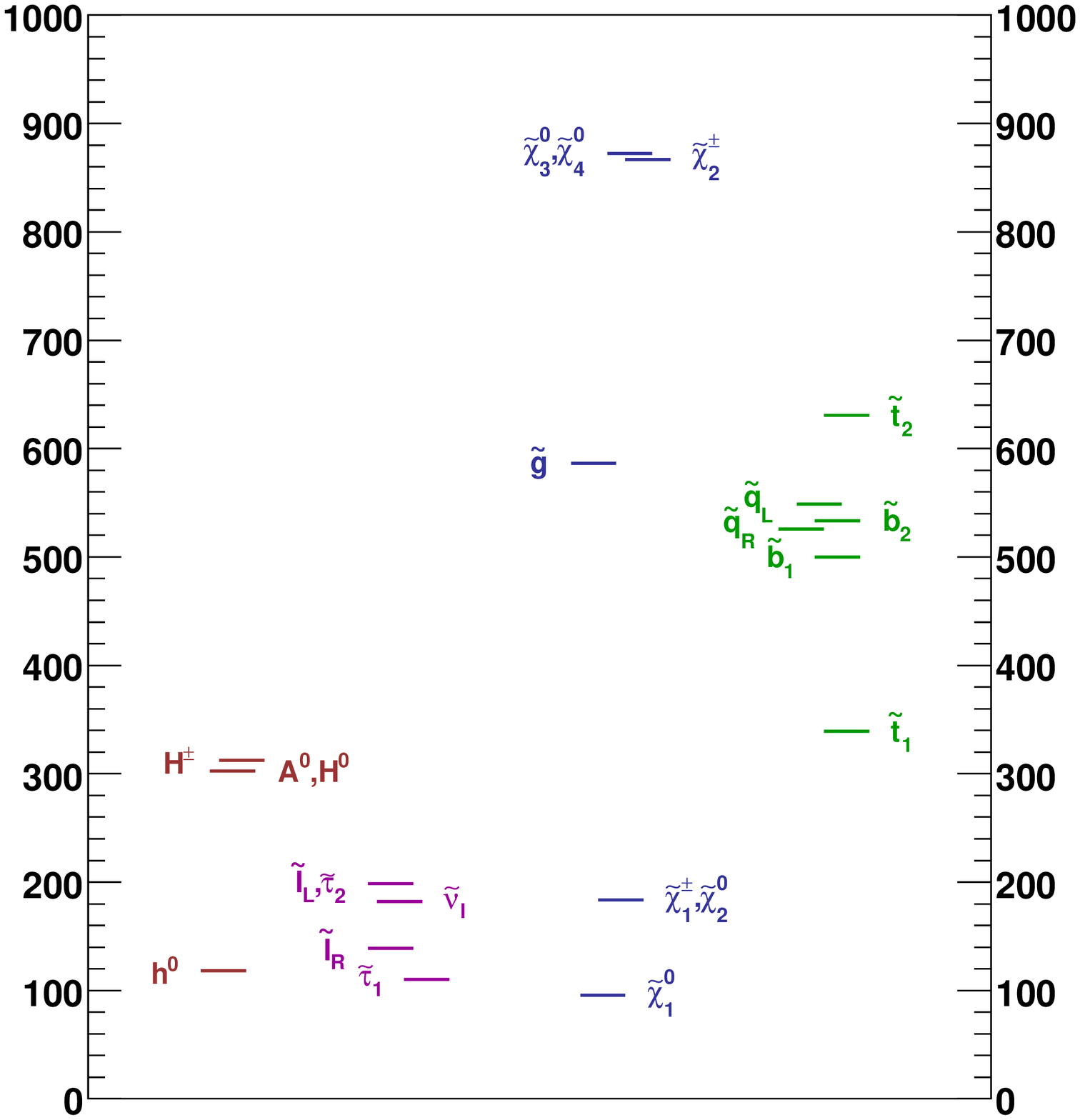}
\caption{SUSY mass spectra of the best fit points of the CMSSM (left)
and the NUHM1 (right)~\cite{mastercode2}.}
\label{fig:bestfit}
\end{center}
\vspace{-1em}
\end{figure}


\subsection*{Acknowledgements}

I thank my collaborators O.~Buchm\"uller, R.~Cavanaugh, A.~De~Roeck,
J.~Ellis, H.~Fl\"acher, G.~Isidori, K.~Olive, P.~Paradisi, F.~Ronga and
G.~Weiglein with whom parts of the results presented here have been
obtained.




\begin{footnotesize}

\end{footnotesize}

\end{document}
